\documentclass[final]{aipproc}
\usepackage{amsmath,amssymb}

\layoutstyle{8x11single}

\begin{document}

\title{Nuclear astrophysics and electron beams}


\author{A.\ Schwenk\footnote{E-mail:~schwenk@physik.tu-darmstadt.de}}{%
address={Institut f\"ur Kernphysik, Technische Universit\"at Darmstadt, 
64289 Darmstadt, Germany and\\
ExtreMe Matter Institute EMMI, GSI Helmholtzzentrum f\"ur
Schwerionenforschung GmbH,\\ 64291 Darmstadt, Germany}}

\begin{abstract}
Electron beams provide important probes and constraints for nuclear
astrophysics. This is especially exciting at energies within the
regime of chiral effective field theory (EFT), which provides a
systematic expansion for nuclear forces and electroweak operators
based on quantum chromodynamics. This talk discusses some recent
highlights and future directions based on chiral EFT, including
nuclear structure and reactions for astrophysics, the neutron skin and
constraints for the properties of neutron-rich matter in neutron stars
and core-collapse supernovae, and the dark matter response of nuclei.
\end{abstract}

\maketitle

\section{Chiral effective field theory}

\begin{figure}[t]
\includegraphics[trim=23mm 18mm 149mm 29mm,width=0.44\textwidth,clip=]%
{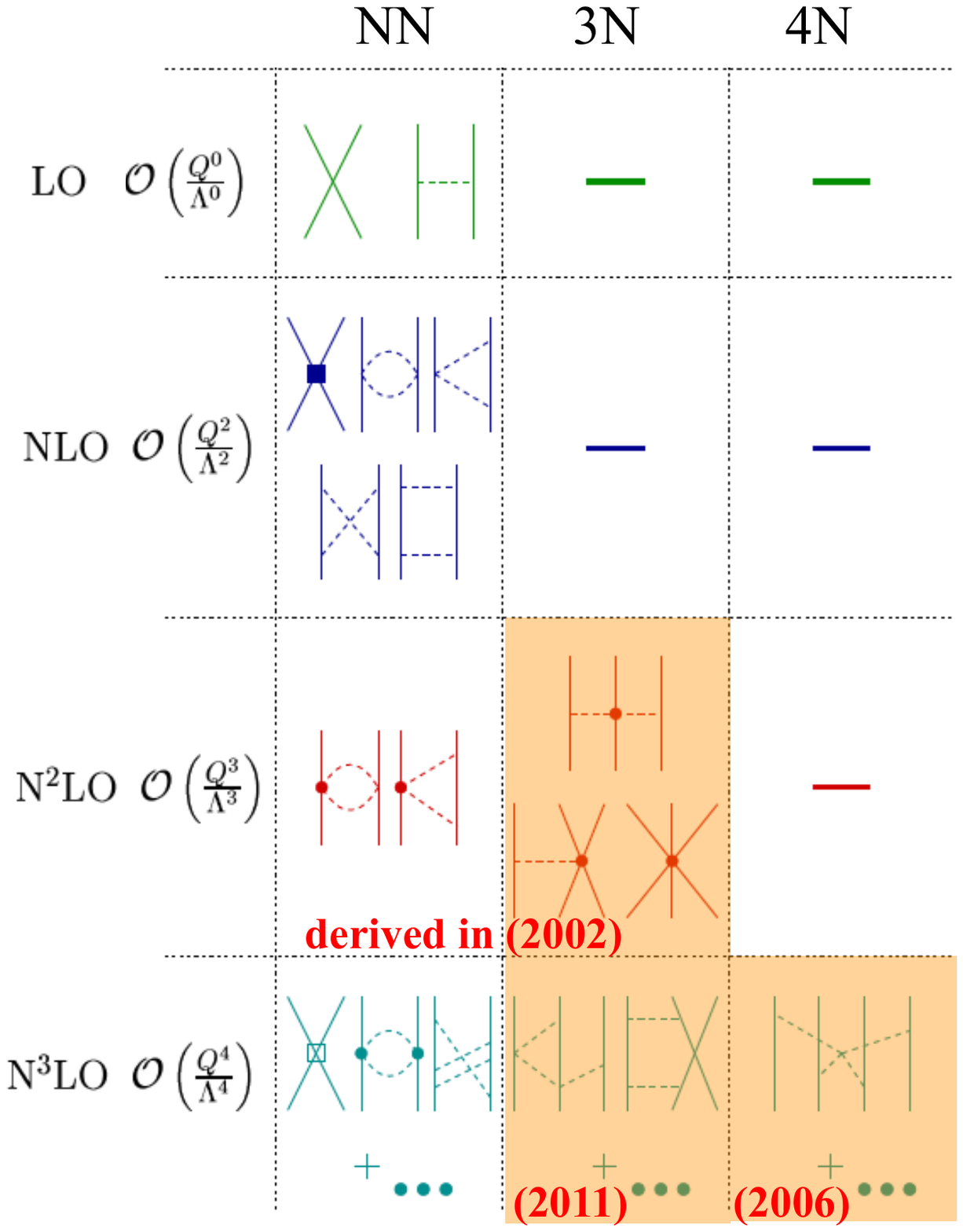}
\hspace*{2mm}
\raisebox{23.75mm}{\includegraphics[scale=0.525,clip=]%
{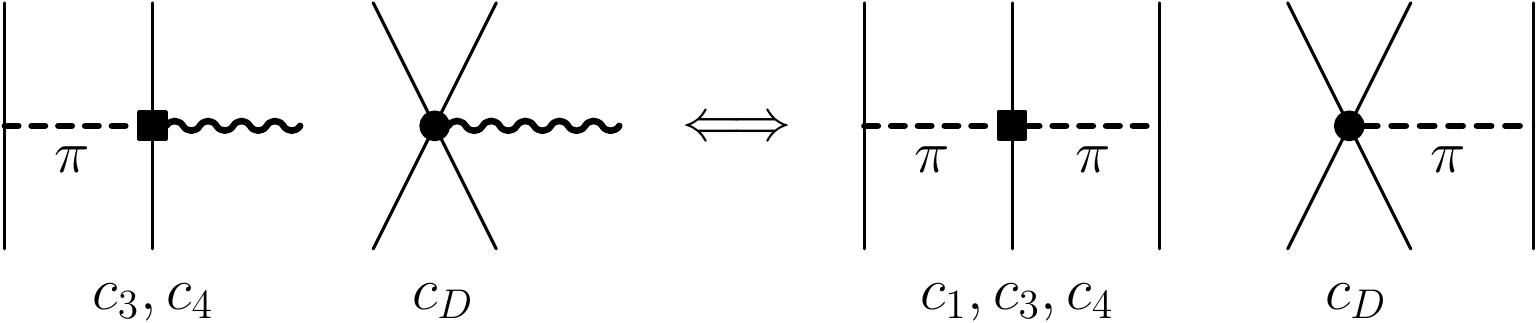}}
\caption{Chiral EFT for nuclear forces, where the different contributions
at successive orders are shown diagrammatically~\cite{RMPEFT,EMRept}. 
Many-body forces are highlighted in orange including the year they
were derived. All N$^3$LO 3N and 4N forces are predicted
parameter-free. On the right, as an example for electroweak operators,
we show the leading two-body axial-vector weak currents that enter at
order $Q^3$ and have the same couplings as in the leading 3N
forces (indicated by the arrow)~\cite{Park}.\label{chiralEFT}}
\end{figure}

Chiral EFT is based on the symmetries of quantum chromodynamics (QCD)
and is applicable at momentum scales of the order of the pion mass $Q
\sim m_\pi$, where pions are included as explicit degrees of freedom
and build up the long-range parts of strong interactions. In chiral
EFT, nucleons interact via pion exchanges and shorter-range contact
interactions~\cite{RMPEFT,EMRept}. The resulting nuclear forces and
consistent electroweak operators are organized in a systematic
expansion in powers of $Q/\Lambda_{\rm b}$, where $\Lambda_{\rm b}
\sim 500 \, {\rm MeV}$ denotes the breakdown scale, leading to a
typical expansion parameter $Q/\Lambda_{\rm b} \sim 1/3$ for nuclei.
As shown in Fig.~\ref{chiralEFT}, at a given order this includes
contributions from one- or multi-pion exchanges and from contact
interactions, with short-range couplings that are fit to data and thus
capture all short-range effects relevant at low energies. The EFT
enables one to estimate theoretical uncertainties and chiral EFT
connects nuclear forces to the underlying theory through lattice
QCD (see talk by W.~Detmold)~\cite{Detmold}.

Chiral EFT opens up a systematic path to investigate many-body forces
and their impact on few- and many-body systems~\cite{RMP3N}. This
results from the consistency of NN and 3N interactions, which predicts
the two-pion-exchange $c_1, c_3, c_4$ parts of 3N forces at N$^2$LO,
leaving only two low-energy couplings $c_D, c_E$ that encode pion
interactions with short-range NN pairs and short-range three-body
physics. The leading N$^2$LO 3N forces improve few-body scattering,
but interesting open problems remain~\cite{Nasser}. This makes the
application of 3N and 4N forces at the next order (N$^3$LO) very
exciting, in particular because they are predicted parameter-free with
many new structures~\cite{RMPEFT}.

Because of gauge symmetries, the same EFT expansion is used to derive
consistent electroweak operators. Therefore, the couplings in nuclear
forces determine also electroweak processes. This provides an
important consistency test and an exciting opportunity for electron
beams. A prime example in chiral EFT are axial-vector weak currents,
where pion interactions (due to the coupling to spin) contribute both
to currents and to nuclear forces. This is already seen at leading
order: The coupling $g_A$ determines the axial one-body current and
the one-pion-exchange potential. Two-body currents, also known as
meson-exchange currents, enter at higher order, just like 3N
forces. As shown in Fig.~\ref{chiralEFT}, the leading two-body
axial-vector currents (at order $Q^3$) are due to long-range
one-pion-exchange and short-range parts~\cite{Park}, with the same
couplings $c_3, c_4$ and $c_D$ of N$^2$LO 3N
forces~\cite{Daniel,Doron}. Chiral EFT is essential for this
connection, which can be viewed as the two-body analogue of the
Goldberger-Treiman relation.

\section{Nuclear structure frontiers}

Recently, nuclear lattice simulations of light $N = Z$ nuclei, based
on lattice chiral EFT interactions, have enabled first ab-initio
calculations of the Hoyle state~\cite{nuclatt}. The triple-alpha
structure and decay of this key state has been studied by electron
scattering at the S-DALINAC~\cite{Hoyle}, which provides precision
tests in the chiral EFT regime.

\begin{figure}[t]
\raisebox{18mm}{\includegraphics[trim=8mm 16mm 75.5mm 88mm,scale=0.55,clip=]%
{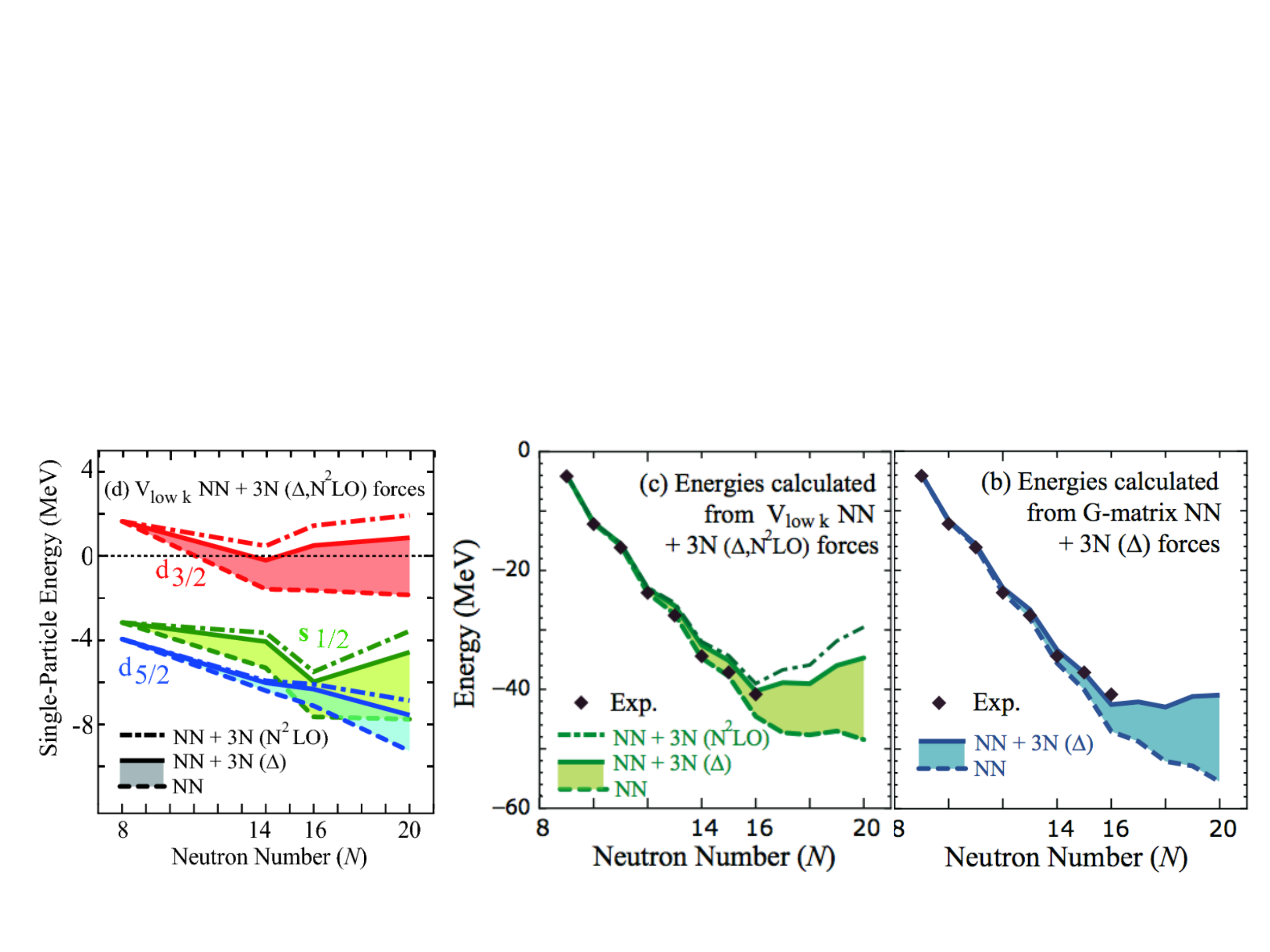}}
\hspace*{2mm}
\includegraphics[trim=0mm 290mm 0mm 0mm,scale=0.4,clip=]{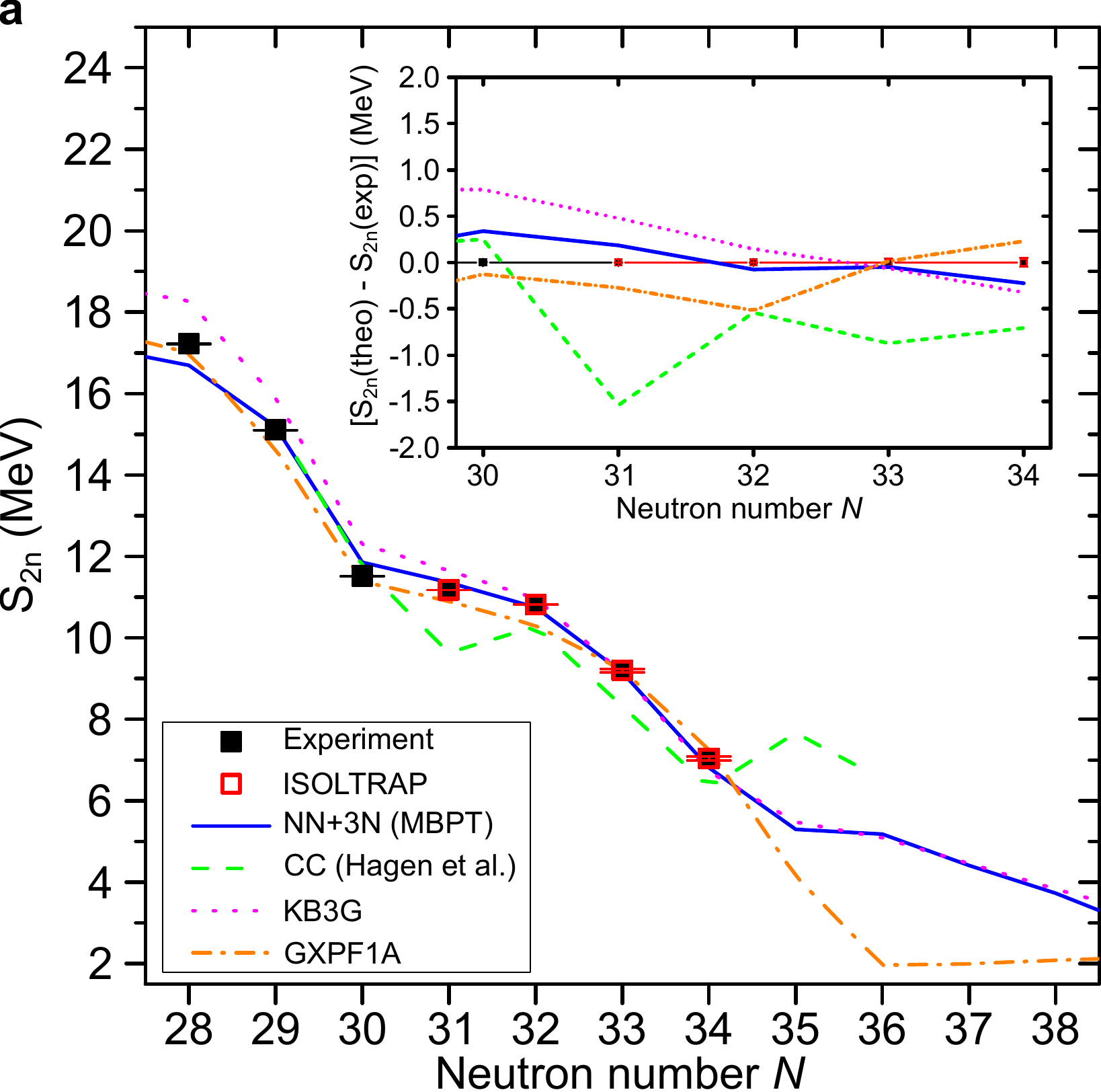}
\caption{Left panel: Single-particle energies in the oxygen isotopes 
as a function of neutron number $N$. Results are shown based on NN
forces only (RG-evolved to low-momentum interactions $V_{{\rm low}\,k}$)
and with N$^2$LO 3N forces (NN+3N). Middle panel: Ground-state
energies of the neutron-rich oxygen isotopes relative to $^{16}$O,
compared to the experimental energies of the bound isotopes
$^{17-24}$O. In both panels, the changes due to the single-$\Delta$
contribution to 3N forces are highlighted by the shaded areas. For
details see Ref.~\cite{Oxygen}. Right panel: Two-neutron separation
energy $S_{2n}$ of the neutron-rich calcium isotopes as a function of
neutron number. Results first published in Nature~\cite{ISOLDE}. The
new ISOLTRAP energies are shown in red. Our predictions based on NN+3N
forces are in excellent agreement with the $S_{2n}$ values, with the flat
behavior from $^{50}$Ca to $^{52}$Ca~\cite{TITAN} and the pronounced
decrease after the $N=32$ shell closure~\cite{ISOLDE} (see also the
inset). For comparison, results are shown for large-space
coupled-cluster (CC) calculations including 3N forces as
density-dependent two-body interactions, and based on the
phenomenological shell-model interactions KB3G and GXPF1A. For details
see Ref.~\cite{ISOLDE}.\label{nrichOCa}}
\end{figure}

Three-nucleon forces are a frontier in the physics of nuclei and for
nucleonic matter in stars. They play a key role in exotic nuclei, for
shell structure and the evolution to the driplines. As shown in the left
panels of Fig.~\ref{nrichOCa}, chiral 3N forces (fit only to $^3$H and
$^4$He) lead to repulsive interactions between valence neutrons that
change the location of the neutron dripline from $^{28}$O (with NN
forces only) to the experimentally observed
$^{24}$O~\cite{Oxygen,Oxygen2}. The position of the neutron dripline
is driven by the location of the $d_{3/2}$ orbital, which remains
unbound with 3N forces. This presents the first explanation of the
oxygen anomaly based on nuclear forces. The 3N-force mechanism is
dominated by the single-$\Delta$ contribution (see the shaded areas in
Fig.~\ref{nrichOCa}) and was recently confirmed in large-space
calculations~\cite{CCO,Heiko,Carlo}.

In studies for calcium isotopes~\cite{Calcium,CCCa,pairing}, it was
shown that 3N forces are key to explain the $N=28$ magic number,
leading to a high $2^+$ excitation energy and concentrated $B(M1)$
transition strength. Our NN+3N predictions for the masses of the
neutron-rich calcium isotopes are shown in the right panel of
Fig.~\ref{nrichOCa}. The predicted flat behavior of the two-neutron
separation energy $S_{2n}$ from $^{50}$Ca to $^{52}$Ca is in
remarkable agreement with precision Penning-trap mass measurements of
$^{51,52}$Ca at TITAN/TRIUMF~\cite{TITAN}, which found $^{52}$Ca to be
$1.74 \, {\rm MeV}$ more bound compared to the atomic mass evaluation.
Recently, the ISOLTRAP collaboration at ISOLDE/CERN was able to
advance the limits of precision mass measurements out to $^{54}$Ca
using a new multi-reflection time-of-flight mass spectrometer. The new
$^{53,54}$Ca masses are in excellent agreement with our predictions
and unambiguously establish $N=32$ as a shell closure (compare the
pronounced decrease in $S_{2n}$ after $N=32$ to after $N=28$), with a
shell gap of almost $4 \, {\rm MeV}$~\cite{ISOLDE}. Moreover, we have
recently extended the NN+3N calculations to the mirror $Z=8, 20$ isotone
chains to study proton-rich exotic nuclei~\cite{protonrich}, which
also test the understanding of isospin-symmetry-breaking nuclear forces.

\section{Neutron skins and neutron stars}

\begin{figure}[t]
\includegraphics[scale=0.315,clip=]{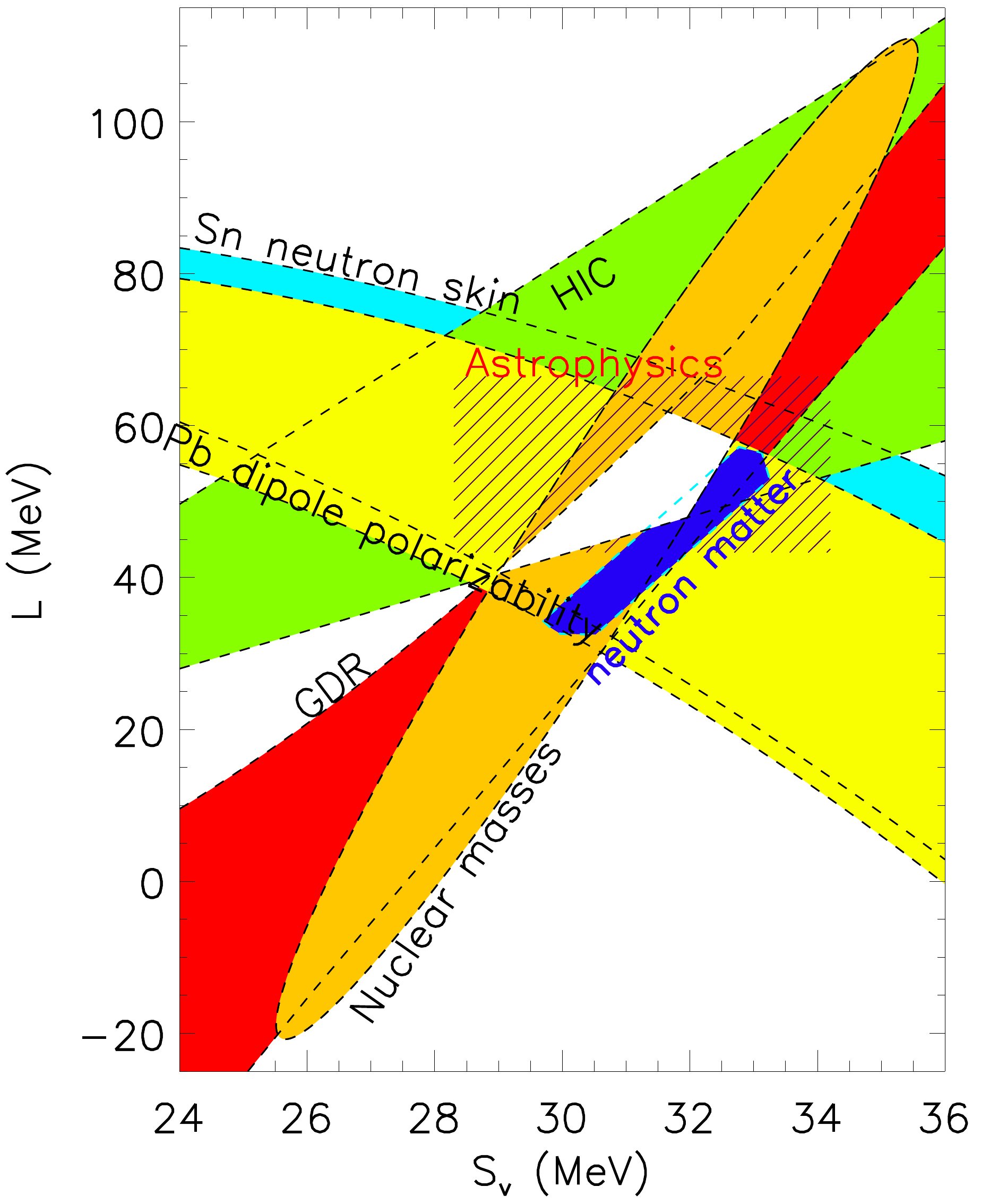}
\hspace*{5mm}
\includegraphics[scale=0.32,clip=]{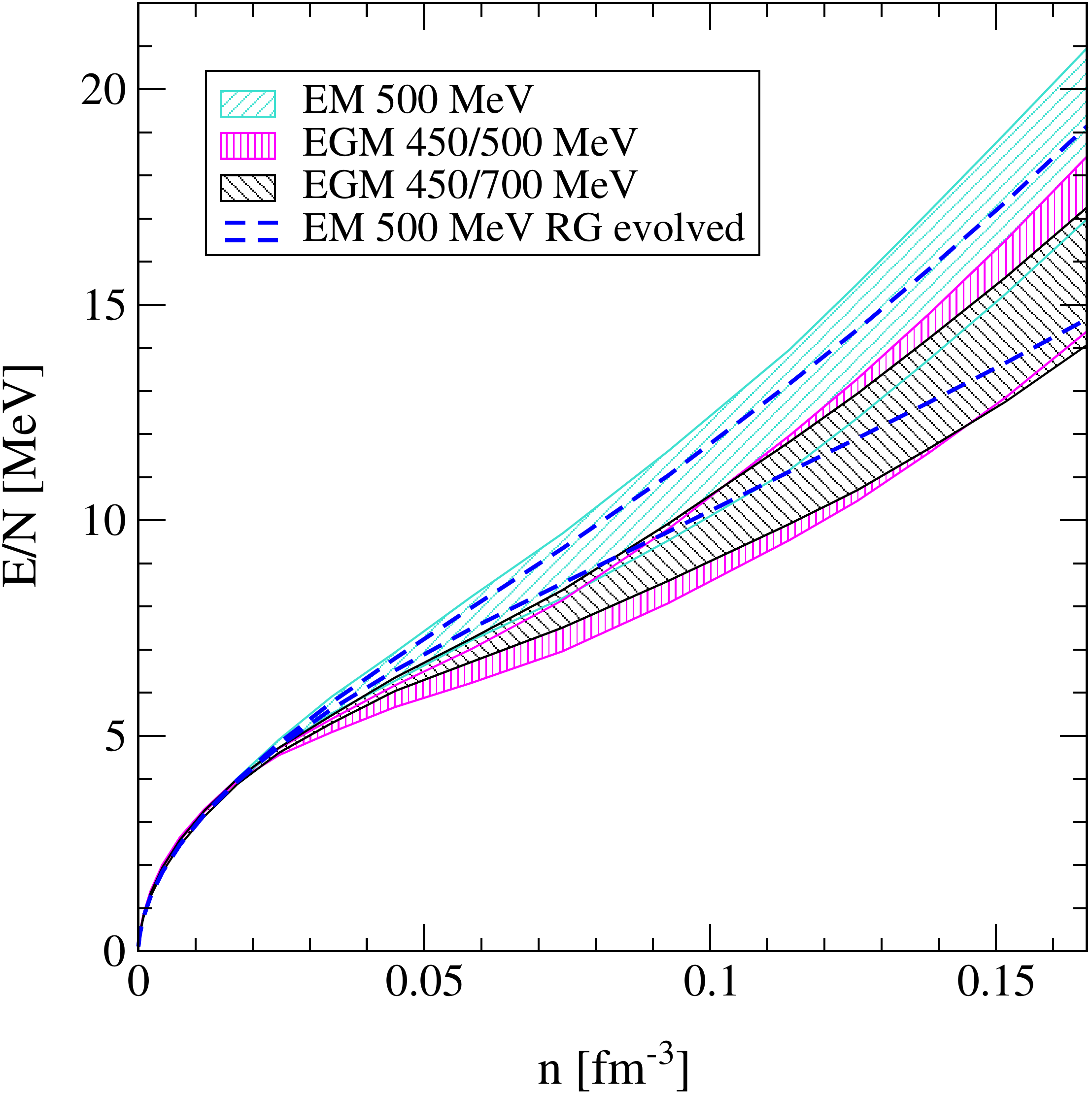}
\caption{Left panel: Constraints for the symmetry energy $S_v$ and
its density derivative $L$. The blue region shows our neutron-matter
constraints~\cite{nstarlong}, in comparison to bands from different
empirical extractions~\cite{LL}. The white area gives the overlap
region of the different empirical ranges. For details see
Ref.~\cite{nstarlong}. Right panel: Neutron-matter energy per particle
$E/N$ as a function of density $n$ including NN, 3N, and 4N forces at
N$^3$LO. The three overlapping bands are based on different NN
potentials (see legend) and include uncertainty estimates due to the
many-body calculation, the low-energy $c_i$ couplings in 3N forces
(which dominate the uncertainty), and by varying the 3N/4N cutoffs.
For comparison, we show the results for the RG-evolved NN EM 500 MeV
potential including only N$^2$LO 3N forces from Ref.~\cite{nm}. For
details see Ref.~\cite{N3LO}.\label{neutmatt}}
\end{figure}

\begin{figure}[!ht]
\includegraphics[scale=0.68,clip=]{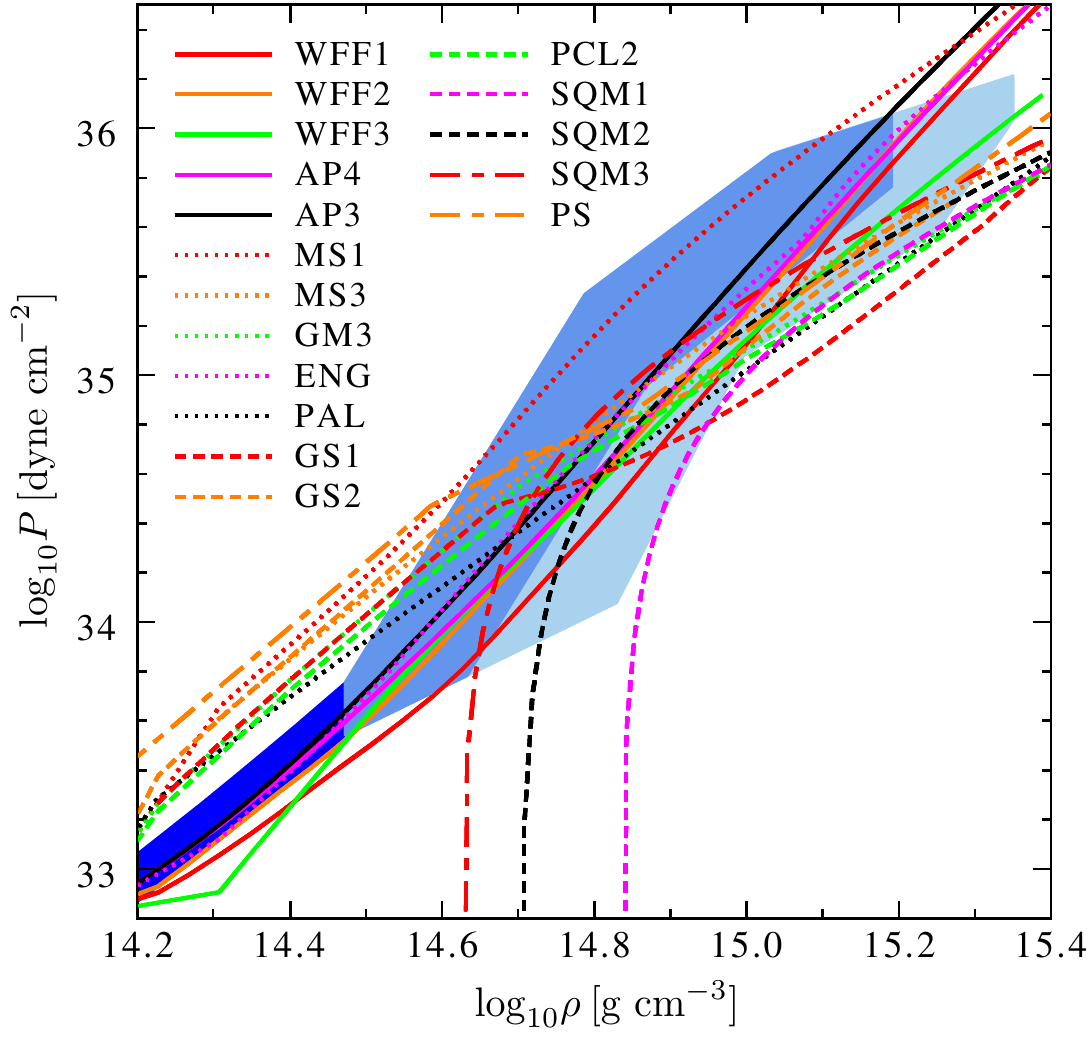}
\hspace*{2mm}
\includegraphics[scale=0.32,clip=]{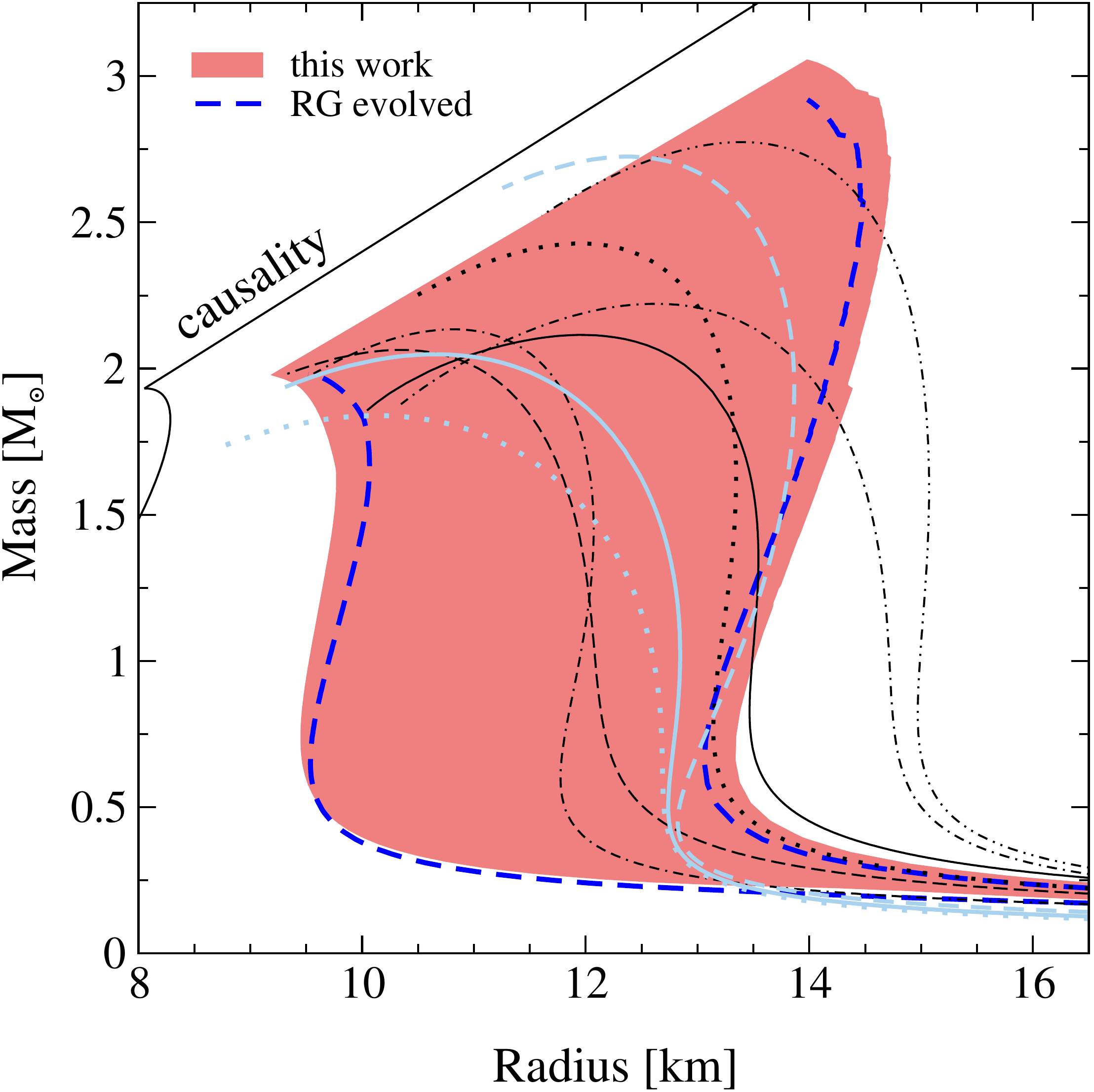}
\caption{Left panel: Constraints for the pressure $P$ of neutron-star
matter as a function of mass density $\rho$ compared to equations of
state commonly used to model neutron stars. The blue band at low
densities represents the pressure predicted by our neutron matter
calculations and incorporating beta equilibrium. The bands at high
densities are the envelope of general polytropic extensions that are
causal and support a neutron star of mass $1.97 \, M_\odot$ (lighter
blue) and $2.4 \, M_\odot$ (darker blue). For details see
Ref.~\cite{nstarlong}. Right panel: Constraints on the mass-radius
diagram of neutron stars based on our N$^3$LO neutron-matter results
and following Ref.~\cite{nstarlong} for the extension to neutron-star
matter and to high densities (red band), in comparison to the
constraints based on RG-evolved interactions (thick dashed blue lines,
based on the $1.97 \, M_\odot$ band of the left panel). We also show
the mass-radius relations obtained from equations of state for
core-collapse supernova simulations. For details see
Ref.~\cite{N3LOlong}.\label{nstar}}
\end{figure}

For systems of only neutrons, the shorter-range $c_D, c_E$ parts of 3N
forces do not contribute because of the Pauli principle and the
coupling of pions to spin~\cite{nm}. Therefore, chiral EFT predicts
all three-neutron and four-neutron forces to N$^3$LO. The same N$^2$LO
3N forces of the previous section are repulsive in neutron matter and
dominate the theoretical uncertainties of the neutron-matter
energy~\cite{nm}. As shown in the left panel of Fig.~\ref{neutmatt},
the predicted energy range provides tight constraints for the symmetry
energy $S_v$ and its density derivative $L$. Neutron skins also probe
the neutron-matter energy and pressure. Our results predict a neutron
skin thickness of $0.17 \pm 0.03 \, {\rm fm}$ for
$^{208}$Pb~\cite{nstar}, in excellent agreement with a recent
determination of $0.156\substack{+0.025 \\ -0.021} \, {\rm fm}$ from
the complete electric dipole response~\cite{Tamii}. In addition, this
can be tested with future determinations of the neutron skin of
$^{208}$Pb and $^{48}$Ca using parity-violating electron scattering at
JLab~\cite{PREX}. See also the talks by J.~Piekarewicz, M.~Dalton, and
C.~Sfienti.

To study the predicted 3N and 4N forces, we have performed the first
complete N$^3$LO calculation of neutron matter including NN, 3N and 4N
interactions~\cite{N3LO,N3LOlong}. The resulting energy is shown in
the right panel of Fig.~\ref{neutmatt}, where the uncertainty band is
dominated by the uncertainty in the $c_i$ couplings of the N$^2$LO 3N
forces. The energy range is also consistent with the RG-evolved
results of Ref.~\cite{nm}. Moreover, first Quantum Monte Carlo
calculations with chiral EFT interactions are providing
nonperturbative benchmarks for neutron matter at nuclear
densities~\cite{Gezerlis}, and there are also calculations of neutron
matter using in-medium chiral perturbation theory
approaches~\cite{Weise,Ulf}.

The neutron-matter calculations based on chiral EFT interactions
constrain the properties of neutron-rich matter below nuclear
densities to a much higher degree than is reflected in current neutron
star modeling~\cite{nstar} and in equations of state for core-collapse
supernova simulations~\cite{N3LOlong}. The constraints for the
pressure of neutron-star matter are shown in the left panel of
Fig.~\ref{nstar} and rule out many model equations of
state~\cite{nstar,nstarlong}.  Combined with the heaviest observed $2
M_\odot$ neutron stars~\cite{Demorest,Antoniadis} and general
extensions to high densities, our N$^3$LO neutron-matter results
constrain the mass-radius of neutron stars to the red region in the
right panel of Fig.~\ref{nstar}, e.g., the radius of a typical $1.4
M_\odot$ star to $R=9.7-13.9 \, {\rm km}$~\cite{nstarlong,N3LOlong}
(the same relative uncertainty as for the neutron skin). The predicted
radius range is due, in about equal amounts, to the uncertainty in 3N
forces and to the extrapolation to high densities. The physics of 3N
forces therefore connects neutron-rich nuclei with neutron stars.  The
radius range is also consistent with astrophysical extractions obtained
from modeling X-ray burst sources (see, e.g., Ref.~\cite{Steiner}).
Finally, the neutron-matter constraints have recently been explored
for the gravitational wave signal in neutron-star mergers~\cite{Bauswein}.

\begin{figure}[t]
\raisebox{2mm}{\includegraphics[scale=0.325,clip=]{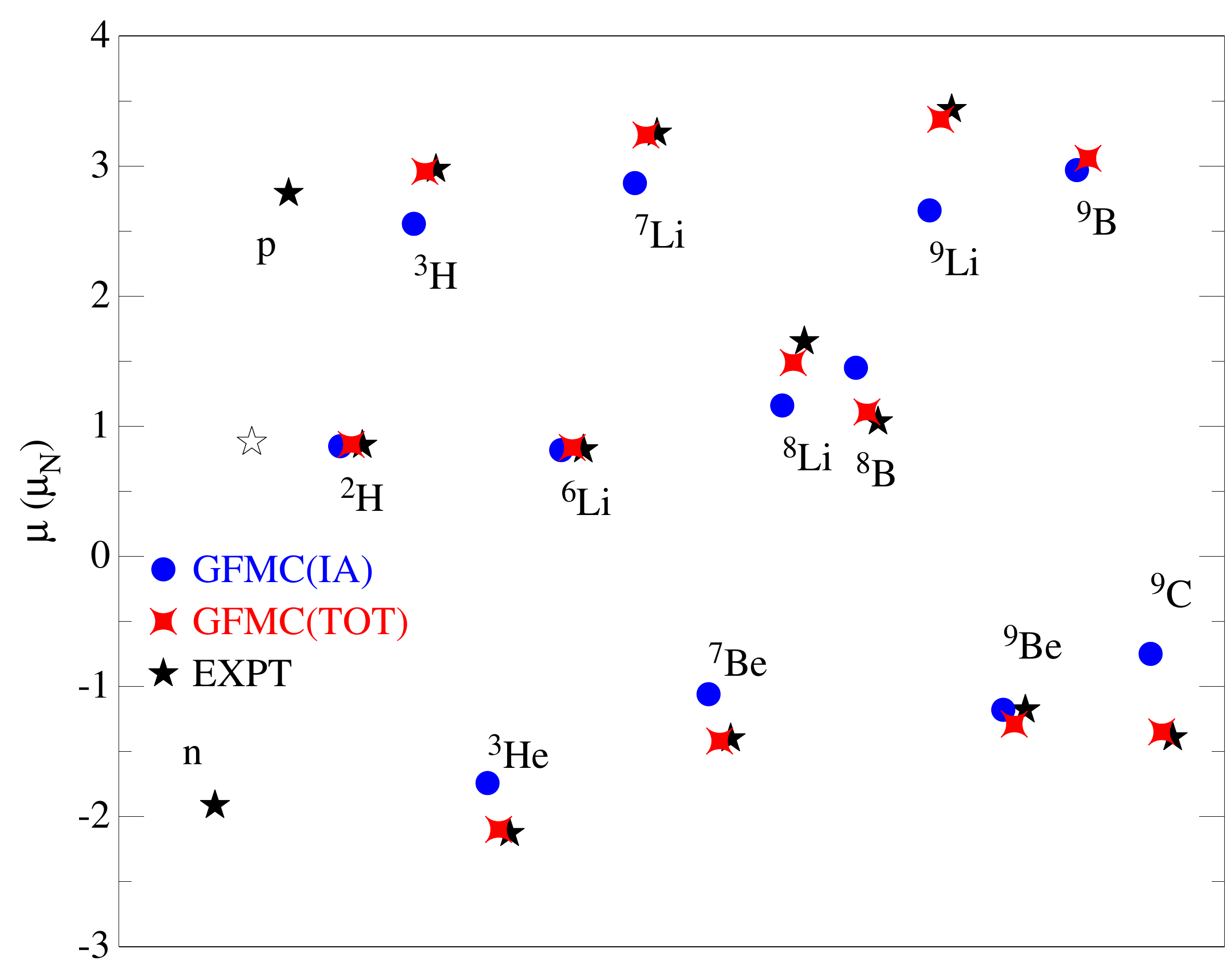}}
\hspace*{5mm}
\includegraphics[scale=0.2475,clip=]{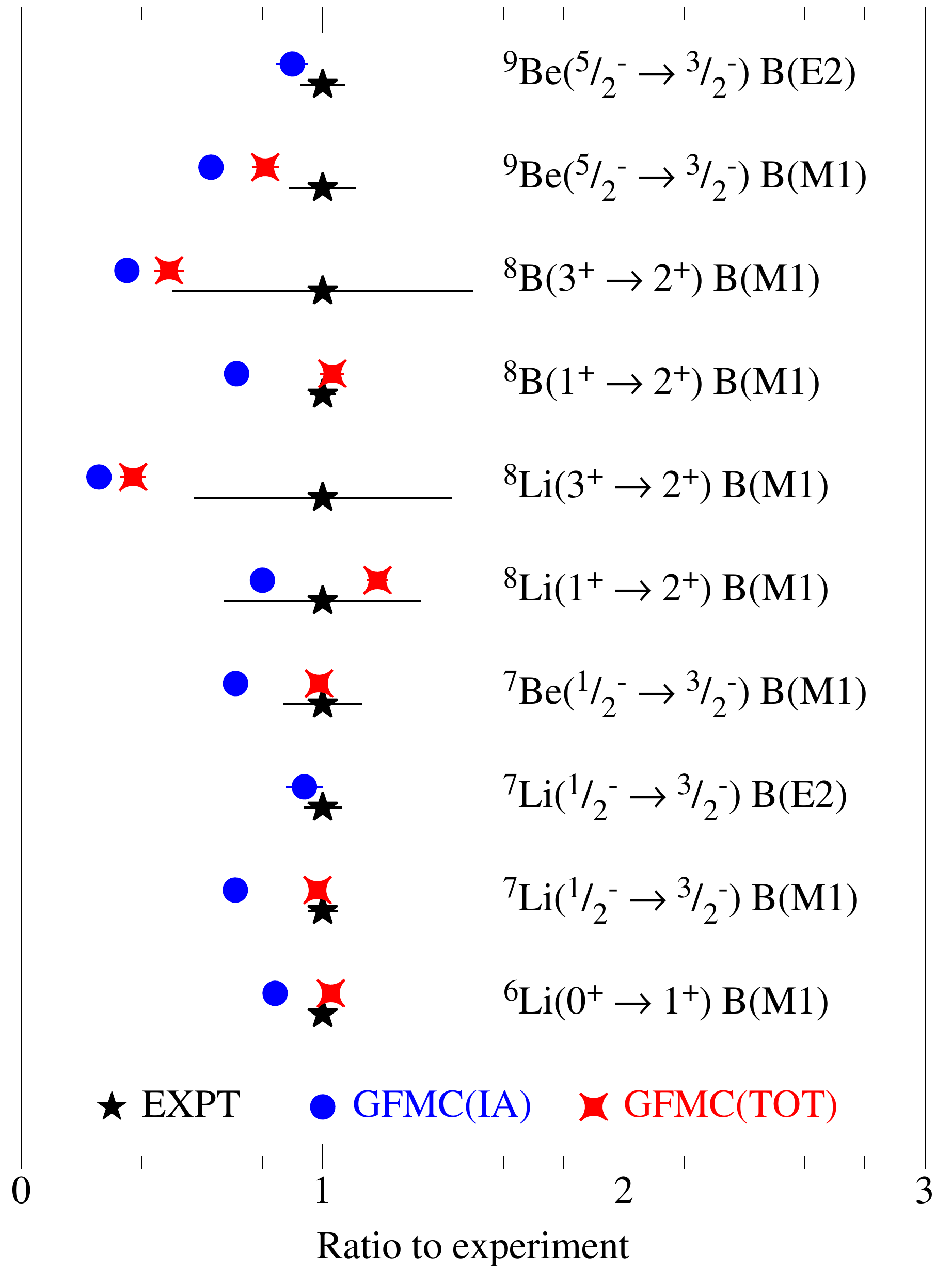}
\caption{Magnetic moments in nuclear magnetons (left panel) and ratio
theory to experiment for $B(M1)$ and $B(E2)$ transition strengths (right panel)
in $A \leqslant 9$ nuclei. Results are based on GFMC calculations
with phenomenological NN and 3N forces, including one-body currents
(IA) and two-body currents to N$^3$LO (TOT) based on chiral EFT, in
comparison with experiment. Figure from Ref.~\cite{Saori}.
Copyright (2013) by The American Physical Society.\label{Saori}}
\end{figure}

\section{Electroweak interactions}

Chiral EFT predicts consistent one- and two-body electroweak currents.
For electromagnetic reactions, two-body currents have been derived
recently and applied to few-nucleon systems~\cite{Kolling,Piarulli}
(see talk by H.~Griesshammer). A highlight are the GFMC calculations
of magnetic moments and electromagnetic transitions in light
nuclei~\cite{Saori} shown in Fig.~\ref{Saori}. The results demonstrate
that two-body currents provide significant contributions to
electromagnetic processes, especially for larger $A$. There are also
interesting sensitivities to 3N forces in electron scattering off
light nuclei~\cite{Sonia}.

While chiral EFT currents have been studied in light nuclei, they were
only recently explored in medium-mass nuclei with a focus on
axial-vector weak currents~\cite{GT0nubb}. Compared to light nuclei,
the contributions of two-body currents are amplified in medium-mass
nuclei because of the larger nucleon momenta. Using a normal-ordering
approximation for two-body currents to create a density-dependent
operator~\cite{Fermi}, it was shown that the leading two-body axial
currents of Fig.~\ref{chiralEFT} contribute mainly to the Gamow-Teller
operator and that the 3N couplings predict a quenching of
low-momentum-transfer Gamow-Teller transitions, dominated by the
single-$\Delta$ contribution~\cite{GT0nubb}. This demonstrates that
chiral two-body currents naturally contribute to the quenching of
Gamow-Teller transitions. A reduction of $g_A$ in the currents is also
expected considering chiral 3N forces as density-dependent two-body
interactions~\cite{JWHolt}.

\begin{figure}[t]
\includegraphics[scale=0.375,clip=]{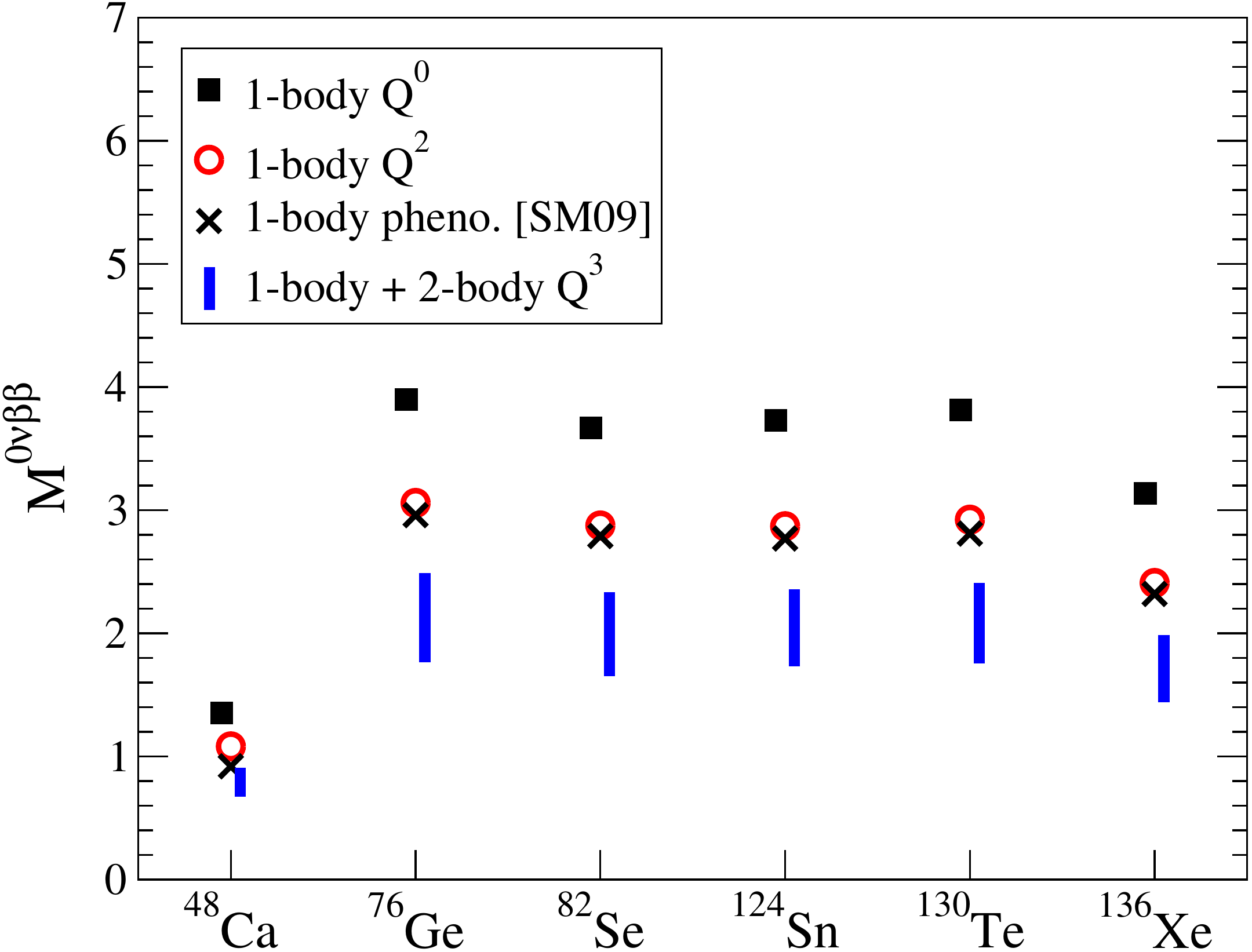}
\caption{Nuclear matrix elements $M^{0\nu\beta\beta}$
for neutrino-less double-beta decay of different nuclei. Results
are shown based on chiral EFT currents at successive orders, including
one-body currents at orders $Q^0$ and $Q^2$, and the predicted 
long-range parts of two-body currents at order $Q^3$
(\cite{GT0nubb}; for details and a discussion of the short-range
contributions, see this reference). For comparison, we also show 
shell-model results (SM09) of Ref.~\cite{0nubbSM} based on
phenomenological one-body currents only.\label{nme}}
\end{figure}

Neutrinoless double-beta decay presents a fundamental test of the
nature of the neutrino, of lepton number, and the neutrino mass scale
and hierarchy. A key input for the ongoing and planned experimental
searches are the nuclear matrix elements that incorporate the
structure of the parent and daughter nuclei and of the decay
mechanism. Compared to standard beta decays, neutrinoless double-beta
decay probes different momentum transfers $Q \sim 100 \, {\rm MeV}
\sim m_\pi$~\cite{0nubbSM}. Therefore, the impact of two-body currents
and renormalization effects can differ from the suppression of $g_A$
in medium-mass nuclei. Chiral EFT predicts the momentum-transfer
dependence of two-body currents, which varies on the order of the pion
mass due to the one-pion-exchange part in the currents in
Fig.~\ref{chiralEFT}. The first calculation of the neutrinoless
double-beta decay operator based on chiral EFT currents at successive
order is shown in Fig.~\ref{nme}~\cite{GT0nubb}. This demonstrates
that the contributions from two-body currents are significant and
should be included in all calculations. An interesting question is how
electron beams can test and constrain the operator and nuclear
structure involved.

A clever idea to study nuclear reactions for astrophysics is to use
electrons to simulate neutrino-nucleus scattering for supernovae and
nucleosynthesis~\cite{simnu}. The relevant inelastic cross sections
are mainly determined by the Gamow-Teller response, which can be
constrained by $(e,e')$ $M1$ data, provided the orbital contribution
is small or can be removed. This has been realized with precision $M1$
data at the S-DALINAC~\cite{simnu}. Recently, there are exciting
developments for nucleosynthesis in core-collapse supernovae due to
observations of ultra metal-poor stars that probe the early chemical
evolution.  Simulations of the neutrino-driven wind in supernovae can
reproduce the lighter r-process Sr to Ag abundances as observed in
ultra metal-poor stars, both for neutron- or proton-rich wind
conditions~\cite{Arcones}. So the reactions for making Sr to Ag are
closer to stability and thus more accessible.

\section{Dark matter response of nuclei}

\begin{figure}[t]
\raisebox{40mm}{\includegraphics[scale=0.35,clip=]{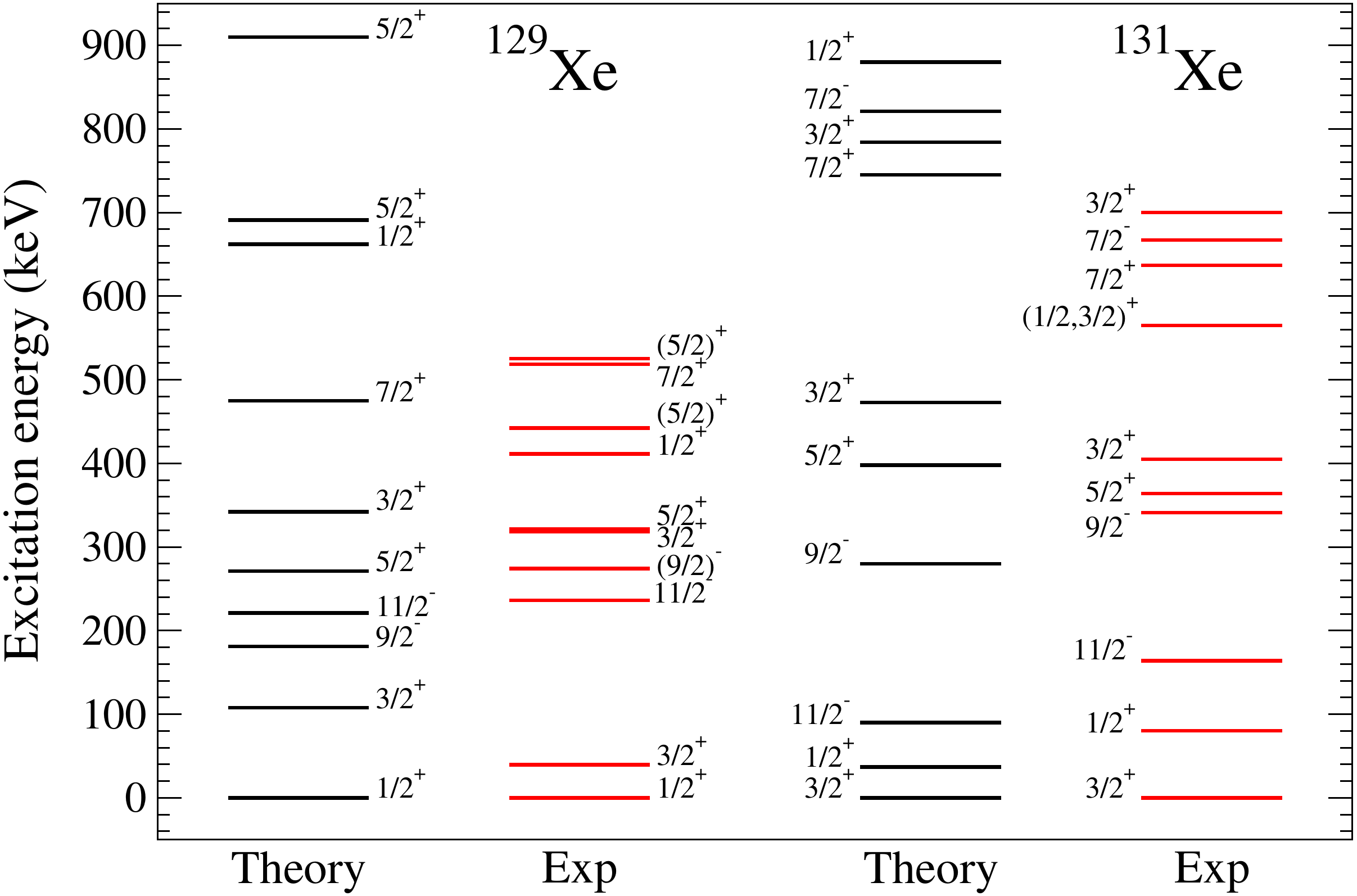}}
\hspace*{2mm}
\includegraphics[scale=0.3,clip=]{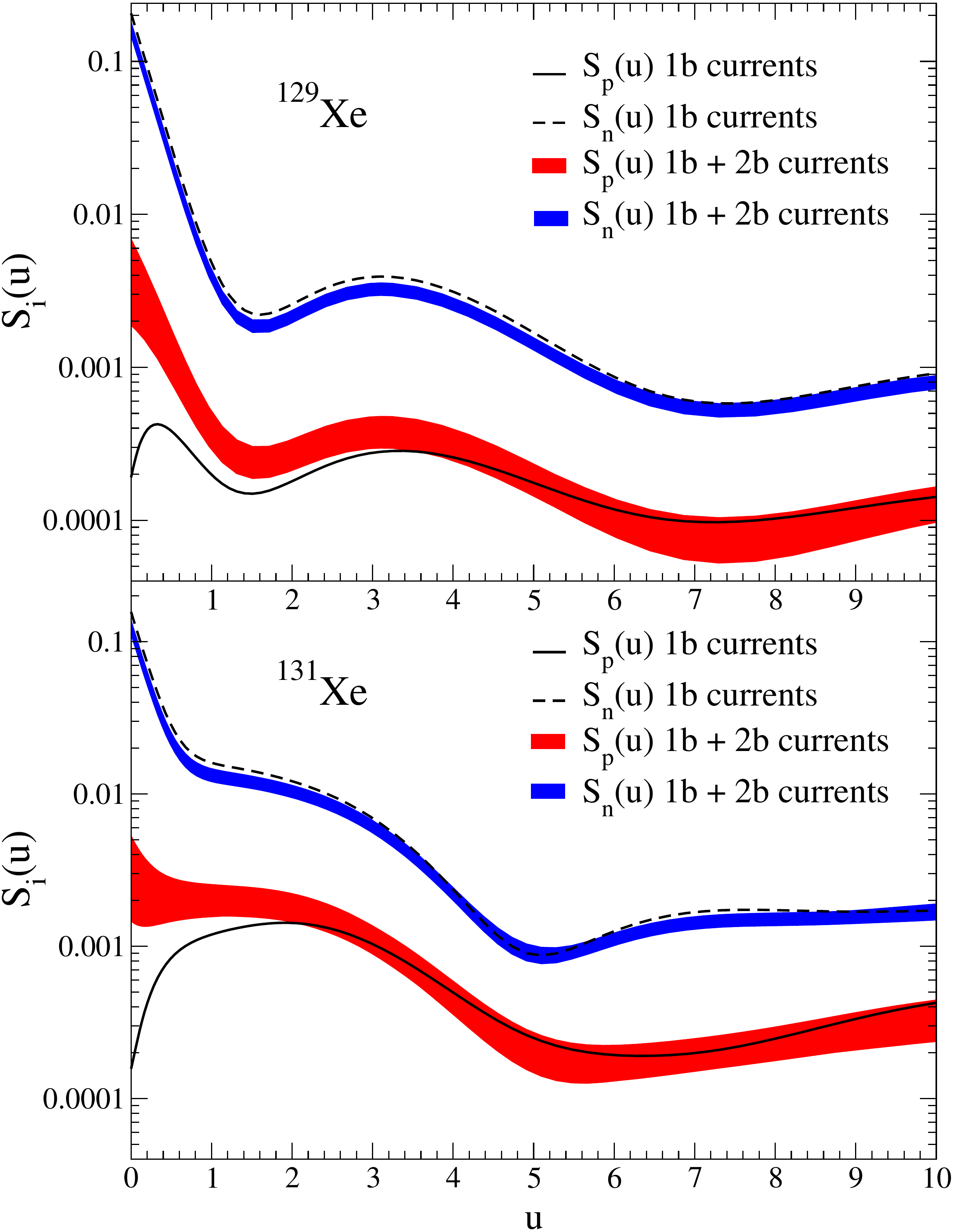}
\caption{Left panel: Comparison of calculated spectra of $^{129}$Xe
and $^{131}$Xe with experiment. For details see Ref.~\cite{DM1}. Right
panel: Structure factors for spin-dependent WIMP scattering off
$^{129}$Xe (top) and $^{131}$Xe (bottom). Results are shown for
WIMP-``proton-only'' $S_{p}(u)$ (solid lines) and ``neutron-only''
couplings $S_{n}(u)$ (dashed lines) as a function of $u=p^2b^2/2$,
with momentum transfer $p$ and harmonic-oscillator length $b$, at the
one-body (1b) current level and including two-body (2b) currents. The
estimated theoretical uncertainty in WIMP-nucleon currents is given by
the red ($S_{p}(u)$) and blue ($S_{n}(u)$) bands. For details see
Refs.~\cite{DM1,DM2}.\label{DM}}
\end{figure}

Direct dark matter detection needs structure factors for elastic
WIMP-nucleus scattering as input, which is particularly sensitive to
nuclear structure for spin-dependent WIMP-nucleon interactions. The
relevant momentum transfers $p$ involved in WIMP scattering off nuclei
are of the order of the pion mass, so that this is a prime regime for
chiral EFT. We have developed the spin-dependent WIMP-nucleon currents
in chiral EFT including the long-range two-body currents, which are
predicted~\cite{DM1,DM2} (see Ref.~\cite{Haxton} for other possible
WIMP-nucleon interactions and Ref.~\cite{Cirigliano} for the
application of chiral EFT to the spin-independent case). The
spin-dependent WIMP-nucleon currents are an isospin rotation of the
axial-vector weak currents shown in Fig.~\ref{chiralEFT}.

We have performed state-of-the-art large-scale shell-model
calculations of the structure factors for spin-dependent WIMP
scattering off $^{129,131}$Xe, $^{127}$I, $^{73}$Ge, $^{19}$F,
$^{23}$Na ,$^{27}$Al, and $^{29}$Si, which covers the non-zero-spin
nuclei relevant to direct dark matter detection~\cite{DM1,DM2}.  For
recent work on the signatures of dark matter scattering inelastically
off nuclei see Ref.~\cite{inelastic}. Our calculations are in the
largest valence spaces with nuclear interactions that have been used
in nuclear structure and decay studies in these mass regions and yield
a good spectroscopic description, as shown for $^{129}$Xe and
$^{131}$Xe in the left panel of Fig.~\ref{DM}. Our results for the
structure factors for spin-dependent WIMP scattering include
theoretical error bands due to the nuclear uncertainties of WIMP
currents in nuclei. The structure factors for $^{129}$Xe and
$^{131}$Xe in the right panel of Fig.~\ref{DM} show that for these
odd-neutron nuclei, two-body currents lead to a significant increase
of the ``proton-only'' structure factors, which follow the
``neutron-only'' ones at low momentum transfers. This is because of
strong interactions between nucleons through two-body currents that
allow the odd species carrying most of the spin to contribute (see
Ref.~\cite{DM2} for a detailed discussion). In fact, with two-body
currents, both ``proton/neutron-only'' structure factors are
determined by the spin distribution of the odd species. Our results
for $^{129}$Xe and $^{131}$Xe were recently used as the benchmark
calculation by the XENON100 collaboration, which provides the best
limits for spin-dependent WIMP-neutron couplings~\cite{XenonSD}. Also
here an interesting question is whether it is possible to
simulate/constrain the dark matter response of nuclei with electron
beams.

\begin{theacknowledgments}
I would like to thank A.~Bauswein, K.~Blaum, J.~Dilling, B.~Friman,
A.~T.~Gallant, D.~Gazit, A.~Gezerlis, K.~Hebeler, J.~D.~Holt,
H.-T.~Janka, P.~Klos, T.~Kr\"uger, J.~M.~Lattimer, J.~Men{\'e}ndez,
C.~J.~Pethick, T.~Otsuka, L.~Schweikhard, J.~Simonis, T.~Suzuki,
I.~Tews, and F.~Wienholtz, who contributed to the results presented in
this talk, and the organizers of this excellent workshop.  This work
was supported by ARCHES, the BMBF under Contract No.~06DA70471, the
DFG through Grant SFB 634, the ERC Grant No.~307986 STRONGINT, and the
Helmholtz Alliance HA216/EMMI.
\end{theacknowledgments}

\bibliographystyle{aipproc}

\end{document}